\documentstyle{article}

\begin{document}

\title{NEAR-EARTH OBSERVATIONS BY SPREAD TELESCOPES}

\maketitle

\date{}

\begin{center}
I.A.\,Maslov,$^1$
S.A.\,Aust,$^1$
V.V.\,Eremin,$^1$
G.I.\,Zubenko,$^1$
A.V.\,Kondabarov,$^1$
O.S.\,Ougolnikov,$^{2,1}$\\

$^{\it 1}$Space Research Institute of RAS, Moscow, Russia,\\
$^{\it 2}$Astro-Space Center, Lebedev's Physical Institute of RAS,
Moscow, Russia
\end{center}

\abstract{
We suggest the all-sky survey at the International
Space Station by four little wide-angle telescopes with
polarization filters and CCD-arrays spread by several meters one
from another. The video information processing will be carried
out by real-time multiprocessor system on the board of the
station. This experiment would allow to observe the sunlit space debris
and meteoroids of centimetre size with their distances and
velocities estimations at the distances up to 20\,km from station
and to investigate the interplanetary and interstellar medium by
the making of polarization sky maps and detecting the
weak-contrast features on it.
}

\section{INTRODUCTION}

The investigations of sky background that basically consists of
zodiacal light are difficult to conduct on the ground because of
sufficient contribution of the light of other sources and
zodiacal light itself (Bernstein {\it et al}, 2002) scattered in
the atmosphere. The translucent high-latitude cirruses found by
IRAS were observed in the visual by Cawson {\it et al} (1986).
Since the scattered light is sufficiently polarized, these
features are better to search for on the polarization sky maps.
The space experiment with polarization sky survey using
wide-angle telescopes and CCD-arrays would allow to investigate
the distribution and properties of interplanetary dust, the size,
shape and orientation of dust particles.

Another problem related with prolonged polarization background
mapping is the possibility of supernova echoes discovering and
investigation. The polarized spots with several years variability
should be observable (Maslov, 2000) around the locations of
supernovas that were observed several centuries ago. Observations
of these spots would give the information about interstellar
medium of our Galaxy. Search for these objects for nonpolarized
light was conducted by Van den Bergh (1966) with photographic
plates and brought the negative result, but using the modern
technique and image processing methods allows to move forward in
this question.

Radar stations of space control are regularly watching for
several thousands spacecrafts and their fragments with sizes more
than 20\,cm. The smallest size particles were registrated by the
collisions with film screens (LDEF satellite) and spacecratfs
surfaces. But experimental data about most dangerous for
spacecrafts medium-sized (about 1\,cm) particles practically are
absent.

\section{EXPERIMENT DESCRIPTION}

The choose of International Space Station (ISS) for this
experiment is made by three reasons:

\begin{itemize}
\item Near-Earth medium watching near the ISS orbit allows to
estimate the statistical parameters and to watch for the bodies
dangerous for the station without additional assumptions about
their space distribution;
\item The size of the ISS allows to spread the telescopes for
triangulational;
\item Installation of the devices on often-visited station
simplifies the changes of the data processing program.
\end{itemize}

The main three goals of the experiment are:

\begin{itemize}
\item Investigations of distribution and properties of dust in
the Solar System and Galaxy;
\item Obtaining the data about space debris and meteoroid particles of
centimetre size near the ISS orbit;
\item Discovering and observations of asteroids, comets and
variable stars.
\end{itemize}

The idea of the experiment follows. Four small telescopes with
$\sim 8^\circ$ field of view installed on ISS, watching at the same
direction and observed the sky by using the continuous station
rotation with the angular velocity about $\sim 4^\circ$/min. The
information processing of CCD-images is being done by on-board
computers in real-time mode.

The light sources coordinates comparison with stars catalogue
allows to determine the exact telescopes orientation, to find
unknown sources and to measure their position. The difference of
these positions measured on different telescopes allows to
determine the distance to the object.

The reasons to use four telescopes are following:

\begin{itemize}

\item using of two or more telescopes spread by several meters
give the possibility to determine the distance to the particle;

\item the registration of the track of fast-moving object by two
telescopes with opposite CCD regimes ("exposition" on the first
and "reading" on the second and vice versa) allows to determine
the angular velocity of the object;

\item using the different axes direction of the polarizing
filters of the telescopes, we can measure the linear polarization
both point and extended sources;

\item the work failure of one telescope does not bring sufficient
change for the worse of the quality of the information remaining
the experiment conduction possible;

\item the measurements exactness is better, the probability of
space debris or meteoroid discovering or unusual event
(short-time burst, for example) observation increases;

\item the extension of observable sky area is possible.

\end{itemize}

The telescopes are installing in pairs on two platforms with
vertical (relatively Earth) axis of rotation. The telescopes are
watching at one direction by the angle about $30-60^\circ$ to the
zenith. The distance between platforms should be not less than
5\,m. The platforms are turning the telescopes to the required
sky region not emitted by the Sun.

\section{APPARATUS DESCRIPTION} 

The apparatus complex consists of two same devices. The mass of
each one is not more than 40\,kg, the size is $940\times
550\times 550$\,mm$^3$, that makes their transfer and
installation on the International Space Station possible.

The telescopes are developed in Space Research Institute basing
on Star Sensor (Ziman, 1994) that is successfully working now on
geostationary communication satellite "Yamal\,100".
The basic parameters of the telescopes are shown in the
Table\,1. The telescopes are able to work at angles down to
$30^\circ$ from the Sun and from the Earth horizon. Adding to the
video information, they supply the parameters of each image
orientation, that is simplify the further information processing.

\begin{table}[h]
\begin{center}
\caption{Telescopes characteristics}
\begin{tabular}{|l|l|}
\hline
Lens diameter & 26\,mm \\
Focal distance & 58\,mm \\
Visible area & $8^\circ \times 8^\circ$ \\
Spectral range & 0.5 -- 1.0\,mkm \\
CCD format & $512\times 512$ \\
Angular resolution & $1^\prime$ \\
Information reading period & 1\,sec \\
Noise of reading & $100 e^-$ \\
\hline
\end{tabular}
\end{center}
\end{table}

The Information Processing and Saving Device is the special board
computer being developed in Space Research Institute. It is
consisted of:

\begin{itemize}
\item two processors of Intel-486 type with the frequency 66\,MHz;
\item energy-independent flash memory not less than 8\,GBit;
\item special modules based on programmed logical matrixes for
fast image processing.
\end{itemize}

One such device can process information from two telescopes. If
we use second device for two telescopes, it will be reserve one
or we will have the possibility of processing programs debugging
and comparison. The basic way to pass the information to Earth
and programs edition is their copying using the ISS server and
changeable information holders.

\section{ALGORYTHMS AND PRINCIPLES OF INFORMATION PROCESSING} 

The input information flux is the sky images made by four
telescopes each second. This flux fulfils the memory of computers
at several minutes, that's why the information compression is
necessary. It is better to do it in order to have the complete
astronomical data (such as maps, catalogues, lightcurves etc.) at
the output. The sources in the images can be classified:

\begin{itemize}

\item by the extension:

\begin{itemize}

\item {\it point sources}: the stars and star-like objects (for
$1^\prime$-resolution);

\item {\it tracks}: the trace as a straight line made by moving object;

\item {\it extended sources}: the source with angular size from
$1^\prime$ to several degrees;

\item {\it background}: the source with angular size more than visible
area.

\end{itemize}

\item by the time averaging:

\begin{itemize}

\item {\it model}: given initially, with parameters correction while
the experiment if necessary;

\item {\it momentary}: present just in one image;

\item {\it current}: present on the map obtained by the images addition
at the single sky survey near the source;

\item {\it seasonal}: present on the map obtained by the images
addition during several months of work (until the data passing to
Earth).

\end{itemize}
\end{itemize}

The tracks and point sources information will be saved in the
catalogs and the extended sources and background - in the
$2^\prime$-resolution sky maps.

Finally, we will obtain the following information:

\begin{itemize}

\item last 200 sky images;

\item point momentary sources catalogue;

\item point current sources catalogue with variability data;

\item current sky map;

\item tracks catalogue;

\item seasonal sky map;

\item current maps of some sky regions;

\item some sky images.

\end{itemize}

The apparatus model is given by "dark image", "flat field image"
and point spread function (PSF) of point source depending on the
orbital declination where the survey was made.

The computer memory is holding the background sky map and stars
catalogue that are the sky model for given spectral region. The
deflections from this model are recording during the survey, that
makes the search for new and variable sources easier. The model
image is the sum of background map and point sources with account
of PSF.

We suggest the following processing sequence:

\begin{itemize}

\item correction of output sky images by "dark image" and "flat
field image";

\item subtraction of model image with the model point sources
brightness correction;

\item photometric calibration by the model stars in the visible
area;

\item the search for tracks and new point sources in images with
subtracted model and their include to the tracks and point
momentary sources catalogues;

\item the creation of current sky map with the size about
$10^\circ \times 10^\circ$ using the images with subtracted
tracks and point momentary sources;

\item the search for weak tracks, stars and extended sources in
the current map;

\item the information about brightness of sky regions out of
current map is including to the seasonal map.

\end{itemize}

The seasonal map of the whole sky with $2^\prime$-resolution
requires about 200-300\,MBytes of memory for each telescope, if
not to take the compression into account. The apparatus
parameters and sky model are being corrected during the time of
experiment and changing after passing the data to Earth.
Polarimetric and parallactic measurements are made basing on the
maps and catalogues obtained by different telescopes.

\section{EXPECTED RESULTS} 

The exactness of single position measurement of an object
relatively the stars is $1^\prime$. Since the stationary and
slowly moving objects are being recorded about 100 times at one
crossing of visible area, the average-squared exactness can reach
$6^{\prime\prime}$. The same exactness can be reached for track
position measurement perpendicular its direction, since this
estimation is made by about 500 pixels.

The sensitivity of the telescopes (by S/N level equal to 1) will
be about $10^m$ for point objects. The exactness of photometric
measurements for bright star-like objects will be about 10
percents. The magnitude of the object present in all images
obtained during 2-minute survey, and magnitude of the track can
be estimated with exactness $0.01-0.02^m$.

The magnitude of space debris or meteoroid with albedo about
0.1 and the size $D$, flying at the distance $r$ with tangential
velocity $v_t$ can be estimated by using Bagrov and
Vygon's (1998) formula:

\begin{equation}
\label{1}
m_*=-31.1+2.5\lg\left(\frac{r^2}{D^2}\frac{v_t}{r}\frac{1}%
{\beta}\right)\mbox{,}
\end{equation}
where $\beta$ is the angular size of one image pixel which is
equal to about $3\times 10^{-4}$. Corresponding to this formula,
fragment with size equal to 1\,cm, flying at 20\,km from ISS with
velocity 40\,km/s will be recorded by the telescopes as a
$10^m$-track, i.e. with S/N ratio equal to 1. With the velocity
or the distance decrease the S/N ratio will rise back
proportional to these parameters.

If we increase the distance between the telescopes to 5-6 meters,
than the parallax of the fragment at the distance equal to 20\,km
will reach $1^\prime$ and it will be possible to measure it with
10-percent uncertainty.

The angular velocity of slow moving (from 0.001 to $1^\circ$/sec)
fragments can be measured by the displacement of the object in
different images. Having measured the angle between the tracks
recorded by CCD-matrixes of two telescopes (in "exposition" and
"reading" regimes) we can determine the angular velocity of fast
moving (from $0.1$ to $8000^\circ$/sec) fragments. The velocity
of meteoroid equal to 40\,km/s can be measured from the distance
300 meters!

Thus, the apparatus will be able to find and measure the
brightness, angular velocity and distance and estimate the size
of all space debris and meteoroids larger than 1\,cm, flying
at the distances from 1 to 20\,km. If the flux of such fragments
is dangerous by possible collisions with the station one time per
10 years, than they will be revealed by this apparatus complex
several times per day.

Polarimetric observations will be conducted by the comparison of
object brightness at four telescopes with different polaroid
axes. For extended objects with size more than $1^\circ$ it is
possible to measure polarized light with the intensity equal to
$10^{-4}$ from the background (Sholomitskii {\it et al}, 1999),
using the large number of pixels. Sky mapping prolonged for the
several years would decrease this value for one more order and to
investigate the detailed features of Galactic background and
zodiacal light variations.

\section{CONCLUSION} 

The experiment would allow to obtain:

\begin{itemize}

\item distribution and scattering parameters of interplanetary
and interstellar dust by the prolonged regular polarimetric sky
mapping;

\item statistical estimations of concentration, velocities and
sizes of space debris and meteoroids near to International Space
Station orbit;

\item data about Novas at early stages before the registration by
ground-based observatories (especially at low angular distances
from the Sun) and statistical characteristics of bursting and
variable stars.

\end{itemize}

\subsection*{Acknowledgements}

The work is supported by Russian Foundation for Basic Research,
grant 00-02-16396.

\end{document}